\documentclass[12pt]{rspublic}

\usepackage{amsbsy}
\usepackage{amssymb}
\usepackage{amsmath}
\usepackage{natbib}

\begin{document}

\title[Entropy entrainment in superfluids]
{Entropy entrainment and dissipation in finite temperature superfluids}

\author{N. Andersson$^1$ and G. L. Comer$^2$}
\affiliation{$^1$ School of Mathematics, University of Southampton, UK\\
$^2$ Department of Physics and Center for Fluids at All Scales, Saint
Louis University, St. Louis, MO, USA}

\label{firstpage}

\maketitle

\voffset 0.5 truein

\def\be{\begin{equation}}
\def\ee{\end{equation}}
\def\bea{\begin{eqnarray}}
\def\eea{\end{eqnarray}}
\def\n{{\rm n}}
\def\x{{\rm x}}
\def\s{{\rm s}}
\def\N{{\rm N}}
\def\S{{\rm S}}
\def\mun{{\mu_\n}}
\def\mus{{\mu_\s}}

\begin{abstract}{Superfluid hydrodynamics; Dissipative mechanisms}
Building on a general variational framework for multi-fluid dynamics, we
discuss finite temperature effects in superfluids.
The main aim is to provide insight into
the modelling of more complex finite temperature superfluid
systems, like the mixed neutron superfluid/proton superconductor
that is expected in the outer core of a neutron star. Our final results
can also (to a certain extent) be used to describe colour-flavour locked quark superconductors
that may be present at the extreme densities in the deep neutron star core.
As a demonstration of the validity of the model,
which is based on treating the excitations in the
system as a massless ``entropy'' fluid, we show that it is formally
equivalent to the traditional two-fluid approach for superfluid Helium.
In particular, we highlight the fact that the entropy entrainment
encodes the ``normal fluid density'' of the traditional approach. We also show how the superfluid
constraint of irrotationality reduces the number of dissipation coefficients
in the system. This analysis provides insight into the more general problem
when vortices are present in the superfluid, and we discuss how the so-called
mutual friction force can be accounted for in our framework. The end product is a
hydrodynamic formalism
for finite temperature effects in a single superfluid condensate. This framework
can readily be extended to more complex situations.
\end{abstract}

\section{Introduction}

Low temperature physics continues to be a vibrant area of research, providing
a number of interesting and exciting challenges. Many of these are associated
with the properties of superfluids/superconductors, either created in the laboratory or
in the cores of mature neutron stars. Basically, matter appears
to have two options when the temperature decreases towards absolute zero.
According to classical physics one would expect the atoms in a liquid to slow
down and come to rest, forming a crystalline structure. It is, however,
possible that quantum effects become relevant before the liquid solidifies
leading to the formation of a superfluid condensate (a quantum liquid). This
will only happen if the interaction between the atoms is attractive and
relatively weak. The
archetypal superfluid system is Helium. It is well established that He$_4$
exhibits superfluidity below $T=2.17$~K. Above this temperature liquid Helium
is accurately described by the Navier-Stokes equations. Below the critical
temperature the modelling of superfluid He$_4$ requires a ``two-fluid''
description \citep{khalatnikov,wilks,putterman}. Two fluid degrees
of freedom are required to explain, in particular, the presence of a second
sound associated with thermal waves in the system.

Phenomenologically, the basic behaviour of superfluid Helium is easy to
understand if one first considers a system at absolute zero temperature. Then
the dynamics is entirely due to the quantum condensate. There exists a single
quantum wavefunction, and the momentum of the flow follows directly from the
gradient of its phase. This immediately implies that the flow is irrotational.
At finite temperatures, one must also account for thermal excitations (like phonons).
That is, not all atoms remain in the ground state. A second dynamical degree of
freedom arises since the excitation gas may drift relative to the atoms. In
the standard two-fluid model, one makes a distinction between a ``normal''
fluid component\footnote{The model obviously assumes that the excitiations can be treated as a ``fluid'', e.g. that
the mean-free path of the phonons is sufficiently short. This may not be the case at very low temperatures.}
and a superfluid part. The associated densities are to a
large extent statistical concepts, as one cannot physically separate the ``normal fluid''
from the ``superfluid'' \citep{llfluid}. It is important to keep this in mind.

The standard two-fluid model for superfluid hydrodynamics derives from the classic work
of London and Tisza. Yet, there exists a number of alternative approaches to the problem. It
has, for example,  been demonstrated that the two-fluid equations
\citep{khalatnikov,wilks,putterman} can be obtained from a ``single fluid''
model, provided that a distinction is made between a background flow and the
sound waves (phonons) in the system \citep{pr83}. Another interesting strategy
is to apply results from extended irreversible thermodynamics
\citep{joubook,mullerbook} to the Helium problem. Extended thermodynamics was
developed to deal with a number of unattractive features of the classic
results, e.g. the infinite propagation speed of thermal signals associated
with Fourier's law. In order to arrive at a causal description, one
introduces additional variables (motivated by the kinetic theory
of gases) leading to a system with richer dynamics. One of these variables
is the heat flux. In the limit of rapid thermal relaxation one retains
Fourier's law, while a slow relaxation leads to the presence of a second
sound. It is easy to show that the latter limit can be used to describe
superfluid systems \citep{gm,mong}.

We now know that many low temperature systems exhibit superfluid properties.
The different phases of He$_3$ have been well studied, both theoretically and
experimentally \citep{voll}, and there is considerable current interest in
atomic Bose-Einstein condensates \citep{pethick}. In fact, the relevance of
superfluid dynamics reaches beyond systems that are accessible in the
laboratory. It is generally expected that neutron stars, which are formed
when massive stars run out of nuclear fuel and collapse following a supernova
explosion, will contain a number of superfluid phases, see \citet{haensel} for a recent account.
This expectation
is natural given the extreme core density (reaching several times the nuclear
saturation density) and low temperature (compared to the nuclear scale of the
Fermi temperatures of the different constituents, about $10^{12}$~K) of these stars. The outer
regions of a typical neutron star will contain a, more or less solid, crust
where the nuclei form a crystalline lattice. As the density increases the
so-called neutron drip is reached. Beyond this point the crust lattice will
coexist with degenerate neutrons. If the temperature is sufficiently low, below
around $10^9$~K, the strong interaction develops an attractive component (which is necessary for the BCS mechanism to operate) and
these neutrons will be superfluid and may flow through the lattice. When the
density reaches nuclear saturation, the crust lattice gives way
to the fluid core of the star. In the outer parts of this core, the neutron
fluid will co-exist with protons, electrons and perhaps muons. At low
temperatures, both neutrons and protons are expected to form condensates.
Hence one is forced to consider the dynamics of a neutron superfluid
coexisting with a proton superconductor and a relativistic gas of
electrons/muons. Modelling an entire star, with all these different phases, is a serious challenge. Yet, the
nature of the outer core is relatively well understood. The physics of the
deep core is much less certain \citep{haensel}. One alternative is that the composition
continues to change as the presence of more massive baryons (hyperons)
becomes energetically favourable. Another possibility is that the ground state of matter
at high density corresponds to a plasma of deconfined quarks and gluons. The
different phases of matter provide a number of different channels for Cooper
pairing, leading to many potential ``superfluid'' components. In order to
develop a moderately realistic model for a neutron star core we need to
improve our understanding of tricky issues concerning hyperon superfluidity
and quark colour superconductors \citep{alford}. Neutron star observations may
provide the only way to constrain our models for this extreme sector of
physics.

The rapid spin-up and subsequent relaxation associated with radio pulsar
glitches \citep{lyne} provides strong, albeit indirect, evidence for neutron star
superfluidity. The standard model for these events is based on, in the first
instance, the pinning of superfluid vortices (e.g. to the crust lattice) which
allows a rotational lag to build up between the superfluid and the part of the
star that spins down electromagnetically, and secondly the sudden unpinning
which transfers angular momentum from one component to the other leading to
the observed spin-change. Key to the modelling of these events is the vortex
pinning and the mutual friction (see \citet{trev1} for a recent discussion)
between the two components in the star.

The modelling of superfluid neutron star
oscillations has also received considerable attention. It is known that
different classes of pulsation modes can be, more or less clearly, associated
with different aspects of the neutron star model. As an example, a superfluid
star has a set of oscillation modes that arise because of the existence of the
second sound \citep{epstein,gm1,lee,cll,ac01}. The hope is that one will be
able to use future observations, e.g. via gravitational waves, to learn more
about the interior composition of the star \citep{astero}. Particularly
interesting in this respect is the possibility that various oscillation modes
may be unstable. The most promising such instability is (according to
current thinking) associated with the inertial r-modes, see
\citet{nk,nareview} for reviews of the relevant literature. The r-mode instability is expected to be active provided
that the gravitation radiation reaction, which drives the
instability, is more efficient than the different damping mechanisms that
suppress the growth of the mode. On the one hand, this is interesting because
it makes the instability window sensitive to the detailed composition of the
star. On the other hand, it makes realistic modelling of the instability
exceedingly difficult. Having said that, progress has been made on
understanding the nature of the r-modes in a superfluid neutron star \citep{linmen,yole,pca},
in particular the role of the vortex mediated mutual friction
damping. We also know that the bulk viscosity associated with hyperons and
deconfined quarks can affect the results significantly \citep{owen,alfbulk}. In
all cases the effects of superfluidity will be considerable.

So far, studies of the dynamics of superfluid neutron stars have almost
exclusively considered the zero temperature problem. This is an obvious
starting point since i) it simplifies the analysis and ii) mature neutron
stars tend to be ``cold'', with core temperature below $10^8$~K.
However, this logic has an obvious flaw. The
critical temperature at which the different phases of matter become
superfluid is density dependent \citep{visc}. For instance, singlet state pairing of
neutrons is expected to be present from just beyond the neutron drip to some
point in the fluid core. For any given stellar temperature there must therefore
exist transition regions where thermal effects play a dominant role. A
detailed model ought to account for these regions. This involves understanding
the dynamical role of the thermal excitations. The aim of the present paper is to take
some steps towards such an understanding. We will demonstrate the close
connection between the variational multi-fluid framework \citep{prix,monster}
that we have previously used to model the outer neutron star core
\citep{pca,prec1,prec2,glitch}, and  the classic
two-fluid model for He$_4$ at finite temperatures
\citep{khalatnikov,wilks,putterman}. This is an important contribution which
 clearly establishes the viability of the variational multi-fluid approach, and lays the foundation for
future applications to problems of astrophysical relevance.

A similar comparison between the corresponding non-dissipative relativistic formulations
has already been carried out by  \citet{cakha}.
They demonstrate how the convective variational
multi-fluid formalism developed by Carter --- see \citet{carter,livrev} for detailed discussions ---
on which our multi-fluid formalism is based,
can be translated into the
model developed by  \citet{khaleb}.
Our analysis provides additional insight into
how thermal excitations should be accounted for, as well as an idea of the
dissipation coefficients that are needed to complete a finite temperature
model. Even though our aim is not to reformulate the modelling of
superfluid Helium, we believe that our discussion should be of some interest also in that context.
The most relevant contributions may be the variational derivation of the hydrodynamic equations
(and the associated use of truly conserved flux quantities) and the analysis of
the superfluid irrotationality constraint. It should also be noted that our formalism is
spiritually close to the extended thermodynamics approach (this point will be
discussed in detail elsewhere \citep{thermo}). This is an interesting
reflection of the universality of conducting multi-fluid models.

Finally, it is worth noting that even though the single particle species
model we consider here is not relevant for the conditions in the outer
neutron star core it may nevertheless be of use for astrophysics modelling. It
could be relevant for a low (but finite) temperature quark core in the
colour-flavour-locked phase (where a single condensate co-exists with a
phonon gas \citep{manuel}). Because of the potential relevance for gravitational-wave astronomy,
studies of the oscillations and instabilities of such a model
would be very interesting. This is a highly relevant problem since observations of these
phenomena may
shed light on the fundamental ground state of matter at extreme densities.

\section{Flux-conservative two-fluid model}

We take as our starting point the flux-conservative multi-fluid framework
developed by \citet{monster}. We consider the simplest system
corresponding to a single particle, heat conducting fluid that can undergo a transition to a superfluid state.
In the
canonical framework, such systems have two degrees of freedom --- the atoms are
distinguished from the massless ``entropy''. In the following, the former will be identified by
a constituent index $\n$, while the latter is represented by $\s$. This
description is different (in spirit) from the standard two-fluid model for
Helium, and it is relevant to investigate how the two models are
related. In particular, we want to understand better the various dissipative
terms that arise when the system is out of equilibrium. That is, we want to be
able to compare our dissipative formalism to the results in the standard
literature \citep{khalatnikov,putterman}. Our hope is that this will
improve our understanding of the role of the thermal excitations. This would
be an important step toward more realistic modelling of the various
condensates that are expected to be present in a neutron star core.

Our flux-conservative model \citep{monster} combines the usual conservation
laws for mass, energy and momentum with the results from a variational
analysis \citep{prix}. The latter is based on using the particle fluxes $n^\x_i$ as the
main variables and deducing the associated chemical potentials $\mu_\x$ and
the conjugate momenta $p_\x^i$\footnote{Each separate fluid is identified by a constituent index. We use x and y to indicate general
components, and the specific components in the two-fluid model under consideration are labelled by n and s, for
particles and entropy.}. The variational analysis defines the canonical momentum associated with each flux, in the usual way.
 However, because of the so-called entrainment effect
each momentum does not have to be parallel to the associated flux. In the case
of a two-component system, with a single species of particle flowing with
$n^\n_i = n v^\n_i$ and a massless entropy with flux $n^\s_i = sv^\s_i$, where $n$ is the particle number density and $s$ represents
the entropy per unit volume,  the
momentum densities are
\be
   \pi_i^\n = n p_i^\n = mn v_i^\n - 2 \alpha w_i^{\n\s} \ ,
\ee
and
\be
   \pi^\s_i = 2 \alpha w_i^{\n\s} \ ,
\ee
where $w_i^{\n\s} = v^\n_i-v_i^\s$ and $\alpha$ is the entrainment coefficient.
To complete the model we need to provide an energy functional $E=E(n,s,w_{\n\s}^2)$, which then determines the
chemical potential and the entrainment coefficient;
\be
\mu_\n = \left. { \partial E \over \partial n } \right|_{s, w_{\n\s}^2} \ , \quad \mu_\s = \left. { \partial E \over \partial s } \right|_{n, w_{\n\s}^2} \ , \quad \mbox{ and} \qquad
\alpha =  \left. { \partial E \over \partial w_{\n\s}^2 } \right|_{n,s} \ .
\ee

These relations highlight one of the main questions considered in this work. We want to understand the role
and physical nature of the entrainment between particles and thermal excitations represented by
the entropy fluid. This is very different from the entrainment that has so far been considered in neutron star models.
In the most commonly studied context, the entrainment between neutrons and protons arises because of
the strong nuclear interaction. Each neutron (say) is endowed with a virtual cloud of protons, leading to
an effective mass different from the bare neutron mass \citep{sauls,cj}. In the dynamical description, this effect is represented by
the entrainment. This mechanism is familiar from low-temperature systems, e.g. He$_3$ where entrainment couples the
two spin populations, and is well explained by Landau Fermi liquid theory \citep{andreev,boru,nicolas,gusak}.
In this context it may seem somewhat unorthodox to consider ``entrainment'' between particles and entropy.
However, such a mechanism arises naturally in the variational model, and it is clearly relevant to
ask whether it  plays an important role. In fact, if we consider the entrainment as altering the effective mass
of a constituent, then it would be very natural for this mechanism to affect also the entropy. The
entropy entrainment would simply represent the inertia associated with the heat flow. In our view,
this interpretation is conceptually quite elegant and we want to understand to what extent it is
useful in practice.

As discussed by \citet{monster}, the associated momentum equations can be
written\footnote{Throughout this paper we use a coordinate basis to represent tensorial relations.
This means that we distinguish between co- and contra-variant objects, $v_i$ and $v^i$, respectively.
Indices, which range from 1 to 3, can be raised and lowered with the (flat space) metric $g_{ij}$, i.e.,
$v_i = g_{ij} v^j$. Derivatives are expressed in terms of the covariant derivative
$\nabla_i$ which is consistent with the metric in the sense that $\nabla_i g_{kl} = 0$.
This formulation of what is, essentially, a fluid dynamics problem may seem somewhat unfamiliar to
some readers, but it has great advantage when we want to discuss the geometric nature of the
different dissipation coefficients. We will then also use the volume form $\epsilon_{ijk}$ which is completely antisymmetric,
and has only one independent component (equal to $\sqrt{g}$ in the present context).}
\be
   f_i^\n = \partial_t \pi_i^\n + \nabla_j (v_\n^j \pi_i^\n +
   D^{\n j}_{\ \ i} ) + n \nabla_i \left( \mu_\n - \frac{1}{2} m v_\n^2
   \right) + \pi_j^\n \nabla_i v_\n^j  \ , \label{eulern}
\ee
and
\be
f_i^\s = \partial_t \pi_i^\s + \nabla_j (v_\s^j \pi_i^\s + D^{\s j}_{\ \ i} )
+ s \nabla_i T + \pi_j^\s \nabla_i v_\s^j \ , \label{eulers}
\ee
where we have used the fact that the temperature follows from $\mu_\s = T$. In these expressions,
$D^\x_{ij}$ represent the viscous stresses while the ``forces'' $f_i^\x$ allow for momentum
transfer between the two components. In the following we will assume that the
system is isolated, which means that $f_i^\n+f_i^\s=0$.

We want to deduce the general form for the dissipative terms in the equations.
To do this we follow the procedure discussed by \citet{monster}, i.e. we
combine the standard conservation laws with the Onsager symmetry principle. In
the present context, when there is no particle creation, mass conservation
leads to
\be
   \partial_t n + \nabla_j (n v_\n^j ) = \Gamma_\n = 0 \ .
\label{continue}\ee
At the same time entropy can increase, so we have
\be
   \partial_t s + \nabla_j (s v_\s^j ) = \Gamma_\s \ge 0 \ .
\ee

From general principles one can show that the energy loss or gain due to
external influences follows from (cf. Eq.~(33) of \citet{monster})
\be
   \varepsilon^\mathrm{ext} = \sum_\x \left[ v_\x^i f_i^\x + D^{\x j}_{\ i}
   \nabla_j v_\x^i + \left( \mu^\x - \frac{1}{2} m_\x v_\x^2 \right)
   \Gamma_\x  \right] \ .
\ee
In the case of an isolated system $\varepsilon^\mathrm{ext}=0$ so the above
relation can be recast as
\be
   T \Gamma_\s = - f_i^\n w_{\n\s}^i - D^j_{\ i} \nabla_j v_\s^i -
   D^{\n j}_{\ i} \nabla_j w_{\n\s}^i \ , \label{TGs}
\ee
where
\be
D_{ij} = D_{ij}^\n + D_{ij}^\s \ .
\ee

The above results are taken, more or less directly, from \citet{monster}. At
this point we recognize a conceptual mistake in our previous analysis. When
identifying the thermodynamical forces and the associated fluxes that are needed
to complete the dissipative model from (\ref{TGs}), we omitted a number of
terms related to $\nabla_j v_\s^i$. As a result, the models discussed by
\citet{monster} are not as general as they could have been. In fact, if we
were to compare our original formulation to the standard dissipative model for superfluid
Helium \citep{khalatnikov,putterman} several bulk viscosity terms would be
missing.

Let us rework, and correct, the analysis of \citet{monster} in the particular
case of two fluids. From (\ref{TGs}) we identify the three thermodynamic
forces  $w_{\n\s}^i$, $\nabla_j v_\s^i$ and $ \nabla_j w_{\n\s}^i$. The
associated fluxes are $- f_i^\n$, $ - D^j_{\ i}$ and $ - D^{\n j}_{\ i}$.
Following the strategy set out by \citet{monster}, the fluxes will be formed
from linear combinations of the forces in such a way that (the
notation here may seem somewhat elaborate, but it is chosen in order to make
the inclusion of additional fluids in the framework straightforward)
\be
- f_i^\n = L_{ij}^{\n\n} w_{\n\s}^j + \tilde{L}^{\n\n}_{ijk} \nabla^j
  w_{\n\s}^k + \tilde{L}^\n_{ijk} \nabla^j v_\s^k \ ,
\ee
\be
- D^\n_{ij} = \tilde{L}_{ijk}^{\n\n} w_{\n\s}^k  + L^{\n\n}_{ijkl}
\nabla^k w_{\n\s}^l+ \tilde{L}^\n_{ijkl} \nabla^k v_\s^l \ ,
\ee
and
\be
- D_{ij} = \tilde{L}_{ijk}^\n w_{\n\s}^k  + \tilde{L}^\n_{ijkl} \nabla^k
w_{\n\s}^l + L_{ijkl} \nabla^k v_\s^l \ .
\ee
In these expressions we have made use of the Onsager symmetry principle.
Limiting the model to the inclusion of quadratic terms in the forces in
(\ref{TGs}), we find that
\be
L^{\n\n}_{ij} = 2 \mathcal{R}^{\n\n} g_{ij} \ , \label{L1}
\ee
\be
\tilde{L}^\n_{ijk} =  \mathcal{S}^\n \epsilon_{ijk} \ ,
\ee
\be
\tilde{L}^{\n\n}_{ijk} =  \mathcal{S}^{\n\n} \epsilon_{ijk} \ ,
\ee
\be
\tilde{L}^\n_{ijkl} = \zeta^\n g_{ij} g_{kl} + \eta^\n \left( g_{ik} g_{jl} +
g_{il} g_{jk} - \frac{2}{3} g_{ij} g_{kl} \right) + \frac{1}{2} \sigma^\n
\epsilon_{ijm} \epsilon^m_{\ \ kl} \ ,
\ee
\be
L^{\n\n}_{ijkl} = \zeta^{\n\n} g_{ij} g_{kl} + \eta^{\n\n} \left( g_{ik}
g_{jl} + g_{il} g_{jk} - \frac{2}{3} g_{ij} g_{kl} \right) + \frac{1}{2}
\sigma^{\n\n} \epsilon_{ijm} \epsilon^m_{\ \ kl} \ ,
\ee
and
\be
L_{ijkl} = \zeta g_{ij} g_{kl} + \eta \left( g_{ik} g_{jl} + g_{il} g_{jk} -
\frac{2}{3} g_{ij} g_{kl} \right) + \frac{1}{2} \sigma \epsilon_{ijm}
\epsilon^m_{\ \ kl} \ . \label{L2}
\ee

We can reduce the number of unspecified dissipation coefficients by noting
that the conservation of total angular momentum requires $D_{ij}$ to be
symmetric, cf. eq (22) of \citet{monster}. This means that we must have
\be
\mathcal{S}^\n = \sigma^\n = \sigma = 0 \ .
\ee
We are then left with a system that has 9 dissipation coefficients;
$\mathcal{R}^{\n\n}$, $\mathcal{S}^{\n\n}$,  $\zeta^\n$,  $\eta^\n$,  $\zeta^{\n\n}$, $\eta^{\n\n}$, $\sigma^{\n\n}$, $\zeta$ and $\eta$.

To conclude the general analysis, let us write down the final expressions for
the dissipative fluxes. To do this we use the decomposition
\be
\nabla_i v^\s_j = \Theta^\s_{ij} + \frac{1}{3} g_{ij} \Theta_\s +
\epsilon_{ijk} W_\s^k
\ee
where we have introduced the expansion
\be
\Theta_\s = \nabla_j v_\s^j \ ,
\ee
the trace-free shear
\be
\Theta^\s_{ij} = \frac{1}{2} \left( \nabla_i v^\s_j + \nabla_j v^\s_i -
\frac{2}{3} g_{ij} \Theta_\s \right) \ ,
\ee
and the ``vorticity''
\be
W_\s^i = \frac{1}{4} \epsilon^{ijk}( \nabla_j v^\s_k - \nabla_k v^\s_j) \ ,
\ee
associated with the entropy flow.
We will use analogous expressions for gradients of the relative velocity. The definition
of the various quantities should be obvious.

We finally arrive at
\be
- f_i^\n = 2 \mathcal{R}^{\n\n} w^{\n\s}_i + 2 \mathcal{S}^{\n\n} W^{\n\s}_i
\ , \label{f1}
\ee
\be
- D_{ij}^\n = \mathcal{S}^{\n\n} \epsilon_{ijk} w_{\n\s}^k + g_{ij}
(\zeta^{\n\n} \Theta_{\n\s} + \zeta^\n \Theta_\s) + 2 \eta^{\n\n}
\Theta^{\n\s}_{ij} + 2 \eta^\n \Theta^\s_{ij} + \sigma^{\n\n} \epsilon_{ijk}
W_{\n\s}^k \ , \label{f2}
\ee
and
\be
- D_{ij} =  g_{ij} (\zeta^{\n} \Theta_{\n\s} + \zeta \Theta_\s)
+ 2 \eta^{\n} \Theta^{\n\s}_{ij} + 2 \eta \Theta^\s_{ij} \ . \label{f3}
\ee

\section{The superfluid constraint}

Let us now
assume that we are considering a superfluid system. For low temperatures and
velocities the fluid described by (\ref{eulern}) should then be irrotational.
To impose this constraint we need to appreciate that it is the momentum that
is quantised in a rotating superfluid, not the particle velocity \citep{prix}.
This means that we require
\be
\epsilon^{klm} \nabla_l p^\n_m = 0 \ .
\label{quantify}\ee
To see how this affects the equations of motion, we rewrite (\ref{eulern}) as
\be
n \partial_t p_i^\n + n \nabla_i \left[ \mu_\n - \frac{m}{2} v_\n^2 + v_\n^j
p_j^\n \right] - n \epsilon_{ijk} v_\n^j (\epsilon^{klm} \nabla_l p^\n_m) =
f_i^\n - \nabla_j D^{\n j}_{\ \ i} \ .
\ee
That is, using \eqref{quantify} we have
\be
\partial_t p_i^\n + \nabla_i \left[ \mu_\n - \frac{m}{2} v_\n^2 + v_\n^j
p_j^\n \right] = \frac{1}{n} \left[ f_i^\n - \nabla_j D^{\n j}_{\ \ i} \right]
\ . \label{euler_sf}
\ee
If we take the curl of this equation we see that the dissipative fluxes must
satisfy
\be
\epsilon^{ijk} \nabla_j \left[\frac{1}{n} \left( f_k^\n - \nabla_l
D^{\n l}_{\ \ k} \right) \right] = 0 \ .
\ee
In other words, we should have
\be
\nabla_i \Phi = \frac{1}{n} \left( f_k^\n - \nabla_l D^{\n l}_{\ \ k} \right)
\label{phi1}
\ee
for some scalar $\Phi$. This constraint ensures that the superfluid remains
irrotational, i.e., there is no generation of turbulence or vorticity.

In order to satisfy this constraint, it is useful to express \eqref{TGs} in
terms of the variables $j^i_{\n \s} = n w^i_{\n \s}$ and $v^i_\s$ rather
than the variables used in the previous section. This means that we have
\be
T \Gamma_\s = - {\cal F}_i^\n j^i_{\n\s} - {\cal D}^\n_{i j}
\nabla^i j^j_{\n \s}  - D_{i j} \nabla^i v^j_\s
\ee
where we have defined
\be
   {\cal F}^\n_i \equiv \frac{1}{n} \left[ f^\n_i - \left(
   \frac{\nabla^j n}{n} \right) D^\n_{j i}\right] \ ,
\ee
and
\be
   {\cal D}^\n_{i j} \equiv \frac{1}{n} D^\n_{i j} \ .
\ee
It follows that \eqref{phi1} becomes
\be
\nabla_i \Phi =   {\cal F}_i^\n - \nabla^j {\cal D}^\n_{j i} \ . \label{phi2}
\ee
Repeating the analysis from the previous section in terms of the new
variables, we see that the thermodynamic fluxes will now be formed from
\be
- {\cal F}_i^\n = {\cal L}_{ij}^{\n\n} j_{\n\s}^j +
           \tilde{\cal L}^{\n\n}_{ijk} \nabla^j j_{\n\s}^k +
           \tilde{\cal L}^\n_{ijk} \nabla^j v_\s^k  \ , \label{forcen}
\ee
\be
- {\cal D}^\n_{ij} = \tilde{\cal L}_{ijk}^{\n\n} j_{\n\s}^k  +
           {\cal L}^{\n\n}_{ijkl} \nabla^k j_{\n\s}^l +
           \tilde{\cal L}^\n_{ijkl} \nabla^k v_\s^l \ , \label{Dn}
\ee
and
\be
- D_{ij} = \tilde{\cal L}_{ijk}^\n j_{\n\s}^k  +
           \tilde{\cal L}^\n_{ijkl} \nabla^k j_{\n\s}^l + {\cal L}_{ijkl}
           \nabla^k v_\s^l  \ .
\ee
Recall that the conservation of total angular momentum requires $D_{ij}$ to be
symmetric.

Let us now consider the constraint \eqref{phi2}. We need
\be
\nabla_i \Phi = \nabla^j \left({\cal L}^{\n \n}_{j i k l} \nabla^k
                j^l_{\n \s} + \tilde{{\cal L}}^\n_{i j k l} \nabla^k
                v^l_\s\right) - ( \mathcal{L}^{\n\n}_{i j } - \nabla^k
                \tilde{\mathcal{L}}^{\n\n}_{i k j} ) j^j_{\n\s}
                - \tilde{\mathcal{L}}^\n_{i j k} \nabla^j v_\s^k \ .
\ee
That is we must have
\be
\tilde{\mathcal{L}}^\n_{i j k} = 0 \ ,
\ee
and
\be
 \mathcal{L}^{\n\n}_{i j } = \nabla^k \tilde{\mathcal{L}}^{\n\n}_{i k j} \ .
\ee
This leaves us with
\be
\nabla_i \Phi = \nabla^j \left({\cal L}^{\n \n}_{j i k l} \nabla^k
                  j^l_{\n \s} + \tilde{{\cal L}}^\n_{i j k l} \nabla^k
                  v^l_\s\right) \ .
\ee
In other words, we must have
\be
\mathcal{L}^{\n\n}_{jikl} = \hat{\zeta}^{\n\n} g_{ji} g_{kl} \ ,
\ee
and
\be
\mathcal{L}^{\n}_{ijkl} = \hat{\zeta}^{\n} g_{ij} g_{kl} \ ,
\ee
which means that
\be
\Phi =  \hat{\zeta}^{\n\n} \nabla_l j^l_{\n \s} + \hat{\zeta}^{\n} \Theta_\s
\label{phinal} \ .
\ee
Finally, it is straightforward (given the results in the previous section) to
show that
\be
- D_{ij} =  g_{ij} (\hat{\zeta}^{\n} \nabla_l j^l_{\n\s} + \zeta \Theta_\s)
+ 2 \eta \Theta^\s_{ij} \ . \label{Dij}
\ee
That is, only \underline{four} dissipation coefficients remain once we  impose the
superfluid constraint.

We want to compare this result to the standard two-fluid model for Helium, e.g. the
results discussed in chapter~9 of \citet{khalatnikov}. In order to do this, we
need to translate our variables into those that are usually considered. In
addition to providing a useful ``sanity check'' on our analysis, this will
give us a direct translation between the various coefficients. This should be
useful for future modelling of superfluid neutron stars. After all, the Helium
dissipation coefficients have been studied in detail both experimentally and
theoretically (mainly through kinetic theory models).

\section{Translation to the orthodox framework}

The relationship between our framework and the traditional non-dissipative
two-fluid model for Helium has already been discussed by  \citet{prix}. To
extend the discussion to the dissipative problem is, as we will now
demonstrate, straightforward.

\subsection{Non-dissipative case}

It is natural to begin by identifying the drift velocity of the quasiparticle
excitations in the two models. After all, this is the variable that leads to
the ``two-fluid'' dynamics. Moreover, since it distinguishes the flow that is
affected by friction it has a natural physical interpretation. In the standard
two-fluid model this velocity, $v_\N^i$, is associated with the ``normal
fluid'' component. In our framework, the excitations are directly associated
with the entropy of the system, which flows with $v_\s^i$. These two
quantities should be the same, and hence we identify
\be
v_\N^i = v_\s^i \ .
\ee

The second fluid component, the ``superfluid'', is usually associated with a
``velocity'' $v_\S^i$. This quantity is directly linked to the gradient of
the phase of the superfluid condensate wave function. This means that it is,
in fact, a rescaled momentum. As discussed by  \citet{prix} we should identify
\be
v_\S^i = \frac{\pi^i_\n}{\rho} = \frac{p_\n^i}{m} \ .
\ee
where $m$ is the atomic mass.
These identifications lead to
\be
\rho v_\S^i = \rho \left[ \left(1 - \varepsilon\right) v_\n^i + \varepsilon
              v_\N^i \right] \ ,
\ee
where $\varepsilon = 2\alpha/\rho$, with $\rho$ the total mass density. We see that the total mass current is
\be
\rho v_\n^i = \frac{\rho}{1 - \varepsilon} v_\S^i -
\frac{\varepsilon \rho}{1 - \varepsilon} v_\N^i \ .
\ee
If we introduce the superfluid and normal fluid densities,
\be
\rho_\S = \frac{\rho}{1 - \varepsilon} \ , \qquad \mbox{ and } \qquad
\rho_\N =  - \frac{\varepsilon \rho}{1 - \varepsilon} \ ,
\ee
we have the usual result;
\be
\rho v_\n^i = \rho_\S v_\S^i  + \rho_\N v_\N^i \ .
\ee
Obviously, it is the case that $\rho = \rho_\S + \rho_\N$. This completes the
translation between the two formalisms. Comparing the two descriptions, it is
clear that the variational approach has identified the natural physical
variables; the average drift velocity of the excitations and the total
momentum flux. Since the system can be ``weighed'' the total density $\rho$
also has a clear interpretation. Moreover, the variational derivation
identifies the truly conserved fluxes, cf. \eqref{continue}.
In contrast, the standard model uses
quantities that only have a statistical meaning \citep{llfluid}. The density
$\rho_\N$ is inferred from the mean drift momentum of the excitations. That
is, there is no ``group'' of excitations that can be identified with this
density. Since the superfluid density $\rho_\S$ is inferred from $\rho_\S =
\rho-\rho_\N$, it is a statistical concept as well.
Furthermore, the two velocities, $v_\N^i$ and $v_\S^i$, are not individually
associated with a conservation law. From a practical point of view,
this is not a problem. The various quantities can be calculated from
microscopic theory and the results are known to compare well to experiments.
At the end of the day, the two descriptions are (as far as applications are concerned)
identical and the preference of one over the other is very much a matter of taste
(or convention). Having said that, we believe that it is easier to adapt the
variational model to more complex systems, e.g. the mixed superfluids that will be present in a neutron star core
[where key general relativistic effects can be naturally incorporated in our framework \citep{livrev}].

The above results show that the entropy entrainment coefficient follows from
the ``normal fluid'' density according to
\be
\alpha = - \frac{\rho_\N}{2} \left( 1 - \frac{\rho_\N}{\rho} \right)^{-1} \ .
\ee
This shows that the entrainment coefficient diverges as the temperature
increases towards the superfluid transition and $\rho_\N \to \rho$. At first sight,
this may seem an unpleasant feature of the model. However, it is simply a
manifestation of the fact that the two fluids must lock together as one passes
through the phase transition. The model remains non-singular as long as
$v_i^\n$ approaches $v_i^\s$ sufficiently fast as the critical temperature is
approached.

Having related the main variables, let us consider the form of the equations
of motion. We  start with the inviscid problem. It is common to work with the
total momentum. Thus we combine (\ref{eulern}) and (\ref{eulers})
to get
\bea
0 &=& f^\n_i + f^\s_i = \partial_t \left(\pi_i^\n + \pi_i^\s\right) +
      \nabla_l \left(v_\n^l \pi^\n_i + v_\s^l \pi_i^\s\right) + n \nabla_i
      \mu_\n + s \nabla_i T \cr
   && - n \nabla_i \left(\frac{1}{2} m v_\n^2 \right) + \pi_l^\n \nabla_i
      v_\n^l + \pi_l^\s \nabla_i v_\s^l \ .
\eea
Here we have
\be
\pi_i^\n + \pi_i^\s = \rho v_i^\n \equiv j_i
\ee
which defines the total momentum density. From the continuity equation \eqref{continue} we see
that
\be
\partial_t \rho + \nabla_i j^i  = 0 \ .
\ee

The pressure $\Psi$ follows from  \citet{monster}
\be
\nabla_i \Psi = n \nabla_i \mu_\n + s \nabla_i T - \alpha \nabla_i w_{\n\s}^2
     \ .
\ee
We also need the relation
\be
v_n^l \pi_i^\n + v_\s^l \pi_i^\s = v^\S_i j^l + v_\N^l j^0_i
\ee
where we have defined
\be
j^0_i = \rho_\N(v_i^\N - v_i^\S) = \pi_i^\s
\ee
and
\be
 \pi_l^\n \nabla_i v_\n^l + \pi_l^\s \nabla_i v_\s^l =  n \nabla_i \left(
\frac{1}{2} m v_\n^2 \right) - 2 \alpha w_l^{\n\s} \nabla_i w^l _{\n\s} \ .
\ee
Putting all the pieces together we have
\be
\partial_t j_i + \nabla_l \left(v_i^\S j^l + v_\N^l j^0_i\right) + \nabla_i
   \Psi = 0 \ . \label{mom1}
\ee

The second equation of motion follows directly from \eqref{euler_sf};
\be
\partial_t v_i^\S + \nabla_i \left( \tilde{\mu}_\S + \frac{1}{2} v_\S^2
\right) = 0
\ee
where we have defined \cite{prix}
\be
 \tilde{\mu}_\S = \frac{1}{m} \mu_\n - \frac{1}{2} \left(v_\n^i -
v_\S^i\right)^2 \ .
\label{mom2}
\ee

The above relations show that our inviscid equations of motion are identical
to the standard ones, cf. \citet{khalatnikov} and \citet{putterman}. The identified
relations between the different variables also provide a direct way to
translate the quantities in the two descriptions. In particular, we have
demonstrated how the ``normal fluid density'' corresponds to the entropy
entrainment in our model. This answers one of our initial questions: We now
understand the role of the entropy entrainment that arises in a natural way
within the variational framework.

\subsection{The dissipative case}

Let us now move on to the dissipative problem. From \eqref{Dij} we immediately
see that in the dissipative case we need to augment (\ref{mom1}) by the
divergence of
\be
D_{ij} = -  g_{ij} \left[\zeta \Theta_\s + \hat{\zeta}^\n \nabla_l \left(n
w_{\n\s}^l\right)\right] - 2 \eta \Theta^\s_{ij} \ . \label{diss1}
\ee
This result should be compared to the dissipative equations in, for example,
\citet{khalatnikov}. In that description, the dissipation in the total
momentum flux follows from the divergence of
\be
\tau_{ij}= - g_{ij} \left[ \zeta_1\nabla_l \left(j^l - \rho v_\N^l\right) +
 \zeta_2 \nabla_l v_\N^l \right] - 2 \eta \Theta^\s_{ij} \ .
\ee
That is,
\be
\tau_{ij}= - g_{ij} \left[ \zeta_1\nabla_l \left(\rho w_{\n\s}^l\right) +
\zeta_2 \Theta_\s \right] - 2 \eta \Theta^\s_{ij} \ .
\ee
First of all we see that the two shear viscosity coefficients are the same.
Secondly, we identify
\be
\zeta = \zeta_2 \qquad , \qquad \hat{\zeta}^\n = m \zeta_1 \ .
\ee
Moving on to the second momentum equation we need the gradient of, cf.
\eqref{phinal},
\be
\frac{1}{m} \Phi = \frac{1}{m} \left[ \hat{\zeta}^{\n\n} \nabla_l \left(n
 w^l_{\n \s}\right) + \hat{\zeta}^{\n} \Theta_\s \right] \ .
\ee
From \citet{khalatnikov} we see that we should compare this to $h$ where
\be
h = - \zeta_3 \nabla_l \left(j^l - \rho v_\N^l\right) - \zeta_4 \nabla_l v_\N^l
\ee
or
\be
h = - \zeta_3 \nabla_l \left(\rho w_{\n\s}^l\right) - \zeta_4 \Theta_\s \ .
\ee
Once we identify
\be
\hat{\zeta}^{\n\n} = m^2 \zeta_3 \qquad , \qquad \hat{\zeta}^\n = m \zeta_4
\ee
we see that the two formulations agree perfectly. Moreover, it is obvious that
$\zeta_1 = \zeta_4$ as required by the Onsager symmetry.

In order to complete the comparison of the two models, we need to comment on
the (perhaps surprising) absence of dissipative heat flux terms in our model.
At first sight, this would seem to be at odds with the traditional
description \citep{khalatnikov} which contain Fourier's law for the heat
conductivity, $q_i = \kappa \nabla_i T$. For consistency, our model requires
$\kappa = 0$, i.e. the thermal conductivity must vanish. Is this an
unattractive feature of our model? In fact, it is not. First of all, it should
be noted that the heat flux is intimately related to the entropy flow. In a
two-component model one does not have the freedom to introduce an
``independent'' heat flux in addition to the massless entropy flux $n_\s^i$,
without at the same time introducing a new dynamical degree of freedom.
Essentially, the model given by \citet{khalatnikov} is a \underline{three} component model
(it certainly identifies three fluxes).
That this makes sense physically is clear from the fact that the thermal
conductivity in Helium arises from the interaction between phonons and rotons
\citep{khalatnikov}, which can drift at different rates. Our two-fluid model would be
a valid representation of the cold regime where the condensate coexists with a single
excitation component (the thermal phonons). It is well-known that,
the thermal conductivity $\kappa$
vanishes in this case. The model is therefore relevant below 0.8~K or so, in the regime where the
phonon dispersion relation is very close to linear.

It is also relevant to comment on the well-known problems associated with
Fourier's law, i.e. the fact that it leads to a non-causal behaviour of
thermal signals. This issue was one of the main motivations for the
development of extended irreversible thermodynamics \citep{joubook,mullerbook}.
A truly sound model for superfluid Helium ought to reflect these developments.
Even though such a model is yet to be formulated, it is clear that our
approach will allow us to make progress in this direction (by introducing an additional
component representing the rotons). This follows
naturally from the discussion of \citet{thermo} where we demonstrate that the relaxation time
associated with the entropy flux in heat conductivity problems is intimately
related to the entrainment.

We have now achieved the main objective of this work. We have demonstrated
that our dissipative two-fluid formulation, with one of the fluids being
associated with the massless entropy flow, reproduces the orthodox model for
superfluid Helium. This comparison is valuable since it enables us to draw experience for
available results for the various dissipation coefficients, e.g. in terms of their effect on sound waves.
It also
demonstrates that it is straightforward to relate the
variational formulation to standard microphysical calculations.
Perhaps, the most practical insight is that our analysis has explicitly
shown that a full variational treatment of Helium requires three, not two,
fluid degrees of freedom.  It remains to be seen how this will impact
the variational formalism that has been much used to model superfluid neutron star dynamics.

\section{Vortices and mutual friction}

The analysis in the previous two sections provides useful insights into the
dynamics of a single component superfluid at finite temperatures. From a
conceptual point of view, it is obviously important to understand how the superfluid
irrotationally constraint simplifies the dynamics of the two-fluid system, i.e.
that the number of dissipation coefficients is
reduced from nine to four. However, the final model may be of rather limited
practical use.

In reality, the superfluid constraint is too severe. A superfluid
can rotate by forming an array of vortices. To describe such a system, we must
revert to the dissipative fluxes \eqref{f1}-\eqref{f3}. However, this more
general description still fails to account for all dissipative channels in
the problem. In particular, it does not easily accomodate the vortex mediated
mutual friction force. In the simplest description \citep{trev1,hv} we expect
a force
\be
f_i^\mathrm{mf} = \mathcal{B}' \rho_\n n_v \epsilon_{ijk} \kappa^j w_{\n\s}^k
+ \mathcal{B} \rho_\n n_v \epsilon_{ijk} \epsilon^{klm} \hat{\kappa}^j
\kappa_l w_m^{\n\s} \ , \label{mf}
\ee
to act on the particles (with a balancing force affecting the excitations).
Here $n_v$ is the vortex area density and $\kappa^i$ represents
the orientiation of the vortices (the hat represents a unit vector) \citep{trev2}. This
force follows after averaging over a locally straight vortex array.

In \citet{monster} we discussed how this force could be accounted for in our
dissipative model. This analysis was not entirely successful. The main reason
for this is that the variational description assumes that the system is
isotropic. This is obviously no longer the case when one introduces an array
of vortices with a preferred direction. This problem can be resolved in
different ways. One can either add an additional ``fluid'' degree of freedom,
representing the averaged vorticity, to the variational discussion (see
\citet{geurst,yamada} for  interesting discussions and \citet{BL} for a
relativistic account). Formally, this may be the most natural approach. In
particular, since the mutual friction then arises as a linear friction
associated with the drift of vortices relative to the excitations.

A more
direct alternative would be to augment the analysis of the dissipative fluxes
with the preferred direction $\hat{\kappa}^j$. This leads to quite a large
number of possible extra dissipative terms. To see this, let us briefly return to
\eqref{L1}-\eqref{L2}.
These relations followed the assumption that the dissipative fluxes must be
linear in the thermodynamical forces. As a result, a two index coefficient
like $L^{\n\n}_{ij}$ can only be constructed out of the metric $g_{ij}$. If
we have an additional vector in the problem, then a number of additional
two-index objects can be written down. We can then have
\be
L^{\n\n}_{ij} = 2 \mathcal{R}^{\n\n} g_{ij} + \mathcal{R}_1 \hat{\kappa}_i
\hat{\kappa}_j + \mathcal{R}_2 \epsilon_{ijk} \hat{\kappa}^k \ .
\ee
The force resulting from this expression can be written
\be
- f_i^\n = L^{\n\n}_{ij} w_{\n\s}^j =  2 \mathcal{R}^{\n\n} w^{\n\s}_i +
\mathcal{R}_1 \hat{\kappa}_i \left(\hat{\kappa}_j w_{\n\s}^j\right) +
\mathcal{R}_2 \epsilon_{ijk} \hat{\kappa}^k w_{\n\s}^j \ .
\ee
In order to compare this to \eqref{mf} we rewrite the latter as
\be
- f_i^\mathrm{mf} = \mathcal{B} \rho n_v \kappa \left[ w^{\n\s}_i - \left(
\hat{\kappa}_j w_{\n\s}^j\right) \hat{\kappa}_i \right] - \mathcal{B}'
\rho_\n n_v \kappa \epsilon_{ijk} \hat{\kappa}^j w_{\n\s}^k \ ,
\ee
and we see that we should identify
\be
2 \mathcal{R}^{\n\n} = - \mathcal{R}_1 = \mathcal{B} \rho n_v \kappa \quad ,
\quad \mbox{and} \quad \mathcal{R}_2 = \mathcal{B}' \rho_\n n_v \kappa \ .
\ee

This provides a simple and natural generalisation of the dissipative framework
discussed in this paper. Of course, it was designed only to account for the
standard form of the vortex mutual friction. It does not in any way provide a
completely general description of a system with vortices. Such a model would
allow a (possibly quite large) number of additional dissipative terms, and would be much more
complicated. This would nevertheless be an interesting problem to consider.
After all, we have not yet accounted for the vortex tension etcetera \citep{bk,mendell,donnelly}.

\section{Discussion}

In this paper we have developed a dissipative two-fluid model, based on
distinguishing the particle flux from a massless entropy flow. Correcting a
conceptual mistake in a previous analysis \citep{monster} we have formulated a
general model for an isotropic system, which requires the determination of
nine dissipation coefficients. We then demonstrated how imposing the constraint
of irrotationality, which is expected for a (pure) superfluid, reduces the complexity
of the problem. The final model is in one-to-one correspondence with the
classic two-fluid model for He$_4$ and we have provided a translation between
the different variables. This comparison highlights the link between the
entropy entrainment in our model and the ``normal fluid density'' in the
standard description (see also \citet{prix}). Finally, we discussed how the presence of vortices in the
superfluid affects the model. In particular, we indicated how one may account
for the vortex mediated mutual friction force. Our final model should be
directly applicable to low temperature, single ``particle species'' systems,
ranging from laboratory systems to astrophysical objects.

There is considerable scope for future developments in this problem area.
First of all, it would be relevant to allow for causal dissipative heat flux
terms, building on the discussion of \citet{thermo}. Secondly, we want to use
the experience gained here to develop finite temperature models for the
different superfluid components expected to be present in a neutron star core.
The final model discussed in this paper, essentially representing
superfluid Helium at low temperatures, may be immediately relevant (in a certain temperature
regime) for a compact star with a colour superconducting quark core
\citep{alford}. Further work is required to formulate a model for the
coexisting neutron superfluid and proton superconductor expected in the outer
core of a neutron star, as well as the neutron superfluid that penetrates such star's elastic inner crust.
Other exotic phases, like hyperon superfluids, may be
even more complex. However, by demonstrating the intimate link between the entropy entrainment
and the thermal excitations, the present analysis has provided a key
ingredient for such models.

\begin{acknowledgements}
NA acknowledges support from STFC via grant number PP/E001025/1. GLC acknowledges partial support from NSF via
grant number PHYS-0855558.
\end{acknowledgements}

\label{lastpage}

\begin{thebibliography}{25}

\bibitem[{{Alford et al}(2008)}]{alfbulk}
 Alford, M. G., Braby, M., \& Schmitt, A., 2008, Bulk viscosity in kaon-condensed color flavor-locked quark matter, J. Phys. G {\bf 35} 115007

\bibitem[{{Alford et al}(2008)}]{alford}
Alford, M. G., Rajagopal, K., Schaefer, T., \& Schmitt, A., 2008,
Color superconductivity in dense quark matter,
Rev. Mod. Phys. {\bf 80} 1455

\bibitem[{{Andersson}(2003)}]{nareview}
Andersson, N., 2003, Gravitational waves from instabilities in relativistic stars, Class. Quantum Grav. {\bf 20} R105

\bibitem[{{Andersson \& Comer}(2001)}]{ac01}
Andersson, N., \& Comer, G. L., 2001, On the dynamics of superfluid neutron star cores, MNRAS {\bf 328},
1129

\bibitem[{{Andersson \& Comer}(2006)}]{monster}
Andersson, N., \& Comer, G. L., 2006, A flux-conservative formalism for convective and dissipative multi-fluid systems, with application to Newtonian superfluid neutron stars, Class. Quantum Grav. {\bf 23}
5505

\bibitem[{{Andersson \& Comer}(2007)}]{livrev}
Andersson, N., \& Comer, G. L., 2007, Relativistic Fluid Dynamics: Physics for Many Different Scales, Living Reviews in Relativity, {\bf 10} no. 1

\bibitem[{{Andersson \& Comer}(2009)}]{thermo}
Andersson, N., \&  Comer, G. L., 2009, {\em Variational multi-fluid dynamics and causal heat conductivity} in preparation

\bibitem[{{Andersson et al}(2005)}]{visc}
Andersson, N., Comer, G. L., \&  Glampedakis, K., 2005, How viscous is a superfluid neutron star core?, Nucl. Phys. A {\bf 763}  212

\bibitem[{{Andersson \& Kokkotas}(1998)}]{astero}
Andersson, N., \& Kokkotas, K. D., 1998, Towards gravitational wave asteroseismology, Mon. Not. R Astro. Soc.  {\bf 299} 1059

\bibitem[{{Andersson \& Kokkotas}(2001)}]{nk}
Andersson, N., \& Kokkotas, K. D., 2001, Gravitational-Wave Instabilities in Rotating Stars, Int. J. Mod. Phys. D {\bf 10} 381

\bibitem[{{Andersson et al}(2006)}]{trev1}
Andersson, N., Sidery, T., \&  Comer, G. L., 2006, Mutual friction in superfluid neutron stars, MNRAS {\bf 368} 162

\bibitem[{{Andreev \& Bashkin}(1975)}]{andreev}
Andreev, A. F. \&  Bashkin, E. P., 1975, Three-velocity hydrodynamics of superfluid solutions,
Zh. Eksp. Teor. Fiz. {\bf 69} 319

\bibitem[{{Bekarevich \& Khalatnikov}(1961)}]{bk}
Bekarevich, I. L.,  \&   Khalatnikov, I. M., 1961, Sov. Phys. JETP {\bf 13}
643

\bibitem[{{Borumand et al}(1996)}]{boru}
Borumand, M., Joynt, R., Kluzniak, W., 1996, Superfluid densities in neutron-star matter, Phys. Rev. C {\bf 54} 2745

\bibitem[{{Carter}(1989)}]{carter}
 Carter, B., 1989, ``Covariant Theory of Conductivity in Ideal Fluid or Solid Media'',
  in A. Anile and M. Choquet-Bruhat, eds., {\em Relativistic Fluid Dynamics
  (Noto, 1987)},  pp. 1--64, Heidelberg: Springer-Verlag

\bibitem[{{Carter \&  Khalatnikov}(1992)}]{cakha}
Carter, B., \& Khalatnikov, I. M., 1992, Equivalence of convective and potential variational derivations of covariant superfluid dynamics, Phys. Rev. D {\bf 45}  4536

\bibitem[{{Carter \& Langlois}(1995)}]{BL}
Carter, B., and Langlois, D., 1995, Kalb-Ramond coupled vortex fibration model for relativistic superfluid dynamics, Nucl. Phys. B {\bf 454}, 402

\bibitem[{{Chamel}(2008)}]{nicolas}
Chamel, N., 2008, Two-fluid models of superfluid neutron star cores, MNRAS {\bf 388} 737

\bibitem[{{Comer \& Joynt}(2003)}]{cj}
Comer., G. L., \& Joynt, R., 2003, Relativistic mean field model for entrainment in general relativistic superfluid neutron stars, Phys. Rev. D {\bf  68} 023002

\bibitem[{{Comer et al}(1999)}]{cll}
Comer, G. L., Langlois, D., \&  Lin, L. M., 1999, Quasinormal modes of general relativistic superfluid neutron stars, Phys. Rev. D {\bf 60},
104025

\bibitem[{{Donnelly}(1991)}]{donnelly}
Donnelly, R. J., 1991,   {\em Quantized vortices in Helium II}
Cambridge: Cambridge Univ. Press

\bibitem[{{Epstein}(1988)}]{epstein}
 Epstein, R. I., 1988, Acoustic properties of neutron stars, Ap. J. {\bf 333}, 880

\bibitem[{{Geurst}(1989)}]{geurst}
Geurst, J. A., 1989, Hydrodynamics of quantum turbulence in He II: Vinen's equation derived from energy and impulse of vortex tangle, Physica B {\bf 154} 327

\bibitem[{{Glampedakis \& Andersson}(2009)}]{glitch}
Glampedakis, K., \&  Andersson, N., 2009, Hydrodynamical Trigger Mechanism for Pulsar Glitches, Phys. Rev. Lett. {\bf 102} 141101

\bibitem[{{Glampedakis et al}(2007)}]{prec1}
Glampedakis, K., Andersson, N., \& Jones, D.I., 2007, Stability of Precessing Superfluid Neutron Stars, Phys. Rev. Lett {\bf 100} 081101

\bibitem[{{Glampedakis et al}(2008)}]{prec2}
Glampedakis, K., Andersson, N., \& Jones, D.I., 2008, Do superfluid instabilities prevent neutron star precession?, Mon. Not. R. Astro. Soc. in press, preprint arXiv:0801.4638


\bibitem[{{Greco \& M\"uller}(1984)}]{gm}
Greco, A., \&  M\"uller, I., 1984, Extended thermodynamics and superfluidity, Arch. Rat. Mech. Anal. {\bf 85} 279


\bibitem[{{Gusakov et al}(2009)}]{gusak}
Gusakov, M. E., Kantor, E. M., \& Haensel, P., 2009, Relativistic entrainment matrix of a superfluid nucleon-hyperon mixture: The zero temperature limit, Phys. Rev. C {\bf 79} 055806

\bibitem[{{Haensel et al}(2007)}]{haensel}
Haensel, P., Potekhin, A. Y., \& Yakovlev, D. G., 2007,
{\em Neutron stars 1: Equation of state and structure} New York: Springer

\bibitem[{{Hall \& Vinen}(1956)}]{hv}
 Hall, H. E., \&  Vinen, W. F., 1956,
 The Rotation of Liquid Helium II. II. The Theory of Mutual Friction in Uniformly Rotating Helium II, Proc. R. Soc. Lond. A {\bf 238} 215

\bibitem[{{Jou et al}(1983)}]{joubook}
Jou, D., Casas-V\'azquez, J., \& Lebon, G., 1983, {\em Extended
irreversible thermodynamics} Berlin: Springer


\bibitem[{{Khalatnikov}(1965)}]{khalatnikov}
Khalatnikov, I. M., 1965, {\em An introduction to the theory of
superfluidity}  New York: W. A. Benjamin, Inc.

\bibitem[{{Khalatnikov \& Lebedev}(1982)}]{khaleb}
Khalatnikov, I. M.,  \&  Lebedev, V. V., 1982, Relativistic hydrodynamics of a superfluid liquid, Phys. Lett. A {\bf 91} 70

\bibitem[{{Landau \& Lifshitz}(1959)}]{llfluid}
Landau, L.,  and Lifshitz, E. M., 1959, {\em Fluid Mechanics}
Oxford: Pergamon

\bibitem[{{Lee}(1995)}]{lee}
Lee, U., 1995, Nonradial oscillations of neutron stars with the superfluid core,
Astron. Astrophys. {\bf 303}, 586

\bibitem[{{Lindblom \& Mendell}(2000)}]{linmen}
Lindblom, L., \& Mendell, G., 2000, r-modes in superfluid neutron stars,
Phys. Rev. D {\bf 61} 104003

\bibitem[{{Lyne et al}(2000)}]{lyne}
Lyne, A. G., Shemar, S. L., \& Graham Smith, F., 2000, Statistical studies of pulsar glitches, MNRAS {\bf 315}, 534

\bibitem[{{Manuel \& Llanes-Estrada}(2007)}]{manuel}
 Manuel, C.,  \& Llanes-Estrada, F.J., 2007, Bulk viscosity in a cold CFL superfluid, JCAP {\bf 8} 1

\bibitem[{{Mendell}(1991a)}]{gm1}
Mendell, G., 1991a, Superfluid Hydrodynamics in Rotating Neutron Stars. I. Nondissipative equations, Ap. J. {\bf 380}, 515

\bibitem[{{Mendell}(1991a)}]{mendell}
Mendell, G., 1991b, Superfluid hydrodynamics in rotating neutron stars. II. Dissipative Effects, Ap. J. {\bf 380}, 530

\bibitem[{{Mongiovi}(1993)}]{mong}
Mongiovi, M. S., 1993, Extended irreversible thermodynamics of liquid helium II,
  Phys. Rev. B {\bf 48} 6276

\bibitem[{{ M\"uller \& Ruggeri}(1993)}]{mullerbook}
M\"uller, I., \&  Ruggeri, T., 1993, {\em Extended thermodynamics}
New York: Springer

\bibitem[{{Nayyar \& Owen}(2006)}]{owen}
Nayyar. M., \& Owen, B. J., 2006, R-modes of accreting hyperon stars as persistent sources of gravitational waves, Phys. Rev. D {\bf 73}  084001

\bibitem[{{Pethick \& Smith}(2002)}]{pethick}
Pethick, C. J.,  \& Smith, H., 2002, {\em Bose-Einstein condensation in
dilute gases}, Cambridge: Cambridge Univ. Press

\bibitem[{{Prix}(2004)}]{prix}
Prix, R., 2004, Variational description of multifluid hydrodynamics: Uncharged fluids, Phys. Rev. D {\bf 69} 043001

\bibitem[{{Prix et al}(2004)}]{pca}
Prix, R., Comer, G. L., \& Andersson, N., 2004, Inertial modes of non-stratified superfluid neutron stars, Mon. Not. R Astro. Soc. {\bf  348}  625

\bibitem[{{Putterman}(1974)}]{putterman}
 Putterman, S. J., 1974, {\em Superfluid hydrodynamics}
Amsterdam: North-Holland

\bibitem[{{Putterman \& Roberts}(1983)}]{pr83}
Putterman, S. J., \& Roberts, P. H., 1983, Classical non-linear waves in dispersive nonlocal media, and the theory of superfluidity, Physica { \bf 117A} 369

\bibitem[{{Sauls}(1989)}]{sauls}
Sauls, J. A., 1989, p. 457-490 in {\it Timing Neutron stars}, ed. H. Ogelman \& E. P. J. van den Heuvel
Dordrecht: Kluwer Acedemic Publishers

\bibitem[{{Sidery et al}(2008)}]{trev2}
 Sidery, T.,  Andersson, N., \&  Comer, G. L., 2008, Waves and instabilities in dissipative rotating superfluid neutron stars, MNRAS {\bf 385} 335

\bibitem[{{Vollhardt \& W\"olfle}(2002)}]{voll}
Vollhardt, D., \&  W\"olfle, P., 2002, {\em The superfluid phases of
Helium 3}, London: Taylor \& Francis

\bibitem[{{Wilks}(1967)}]{wilks}
Wilks, J., 1967, {\em Liquid and solid Helium} Oxford: Clarendon Press

\bibitem[{{Yamada et al}(2007)}]{yamada}
Yamada, K., Miyake, K., \&  Kashiwamura, K., 2007, Three-Fluid Hydrodynamics for Vortex Turbulence in Superfluid $^4$He,  J. Phys. Soc. Japan
{\bf 76} 014601

\bibitem[{{Yoshida \& Lee}(2003)}]{yole}
Yoshida, S., \& Lee, U., 2003, r-modes in relativistic superfluid stars, Phys. Rev. D  {\bf 67} 124019

\end{thebibliography}
\end{document}